\documentstyle[prc,aps]{revtex}
\input epsfig.sty
\newcommand{\bsigma}{\mbox{\boldmath $\sigma$}}
\newcommand{\btau}{\mbox{\boldmath $\tau$}}

\begin{document}


\title{One Body Density Matrix, Natural Orbits and Quasi Hole States 
in $^{16}$O and $^{40}$Ca}
\author{A. Fabrocini$^{1)}$ and G.Co'$^{\,2)}$}  
\address{
$^1)$ Dipartimento di Fisica, Universit\`a di Pisa,\\ 
and Istituto Nazionale di Fisica Nucleare, sezione di Pisa,\\
I-56100 Pisa, Italy \protect\\
$^2)$
Dipartimento di Fisica, Universit\`a di Lecce \\
 and Istituto Nazionale di
Fisica Nucleare, sezione di Lecce,  \\  
I-73100 Lecce, Italy }
\maketitle

\date{\today}

\begin{abstract}

The one body density matrix, momentum distribution, 
natural orbits and quasi hole states  of $^{16}$O and $^{40}$Ca 
are analyzed in the framework of the correlated basis function theory
using state dependent correlations 
with central and tensor components. 
Fermi hypernetted chain integral equations and single operator 
chain approximation are employed to sum cluster diagrams at all orders. 
The optimal trial wave function is determined by means of the
variational  
principle and the realistic  Argonne $v'_{8}$ 
two--nucleon and Urbana IX  three--nucleon interactions. 
The correlated $^{16}$O momentum distribution 
is in good agreement with the variational Monte Carlo results 
and shows the well known enhancement at large momentum values 
with respect to the independent particle model. 
A similar behavior is found in $^{40}$Ca. The relative importance 
of the different types of correlations (mainly Jastrow and tensor) 
on the momentum distribution appears to be similar in the nuclei 
and in nuclear matter. Diagonalization of the density matrix provides 
the natural orbits and their occupation numbers. 
Correlations deplete the occupation number of the first natural 
orbitals by more than 10$\%$.  
The first following ones result instead occupied by a few percent. 
The single particle overlap functions and the spectroscopic 
factors are computed in the correlated model for both 
nuclei and compared with previous estimates. Jastrow correlations 
lower the spectroscopic factors of the valence states 
by a few percent ($\sim 1-3\%$) with respect to unity. 
An additional $\sim 8-12\%$ depletion is provided by spin--isospin 
 tensor correlations. It is confirmed that a variational teatment of  
short range correlations does not explain the spectroscopic factors 
extracted from $(e,e'p)$ experiments. Such an approach corresponds 
to the zeroth order of the correlated basis function theory and  
two--hole one--particle perturbative corrections in the correlated basis 
are expected to provide most of the remaining strength, 
as in nuclear matter.

\end{abstract}
\vskip 1.cm

%
%

\section{Introduction}

The notion of nuclei as a set of mutually, strongly  interacting particles 
is now generally accepted and it is widely recognized that 
correlations beyond the mean field play a substantial, if not decisive, 
role in the microscopic description of nuclear properties. This 
is intuitive when energies are evaluated, but there are also clear 
signatures of the presence of correlations
in some quantities related to the behavior  
of the single nucleon in the medium. For instance, it is known that  
the one nucleon 
momentum distribution (MD) has dominant high momentum components which are 
due to short range nucleon--nucleon (NN) correlations and are not 
describable in any independent particle model (IPM)\cite{zab78,ant88}. 
In addition,  NN correlations 
may be responsible for the reduction of the spectroscopic strengths 
of the hole states\cite{ste91}. 

The ideal way of describing the nucleus would consist in solving 
the many--body Schr\"odinger equation with realistic interactions. 
Exact solutions have been obtained in light nuclei up to A=8 
within a variety of methods: quantum Monte Carlo\cite{wir00,pud97}, 
Faddeev--like\cite{che86} and correlated hyperspherical 
harmonics\cite{kie93} expansions. 


The treatment of heavier nuclei has not yet attained the same degree 
of accuracy as the light ones. However, the situation is rapidly 
improving, at least for doubly closed shell nuclei. 
The $^{16}$O nucleus has been studied by 
variational Monte Carlo\cite{pie92} (VMC) 
and coupled cluster\cite{hei99} methods.
The variational theory underlying the correlated basis function 
(CBF)\cite{wir88} method has been used 
in doubly closed shell nuclei by Fermi hypernetted chain (FHNC) 
integral equations\cite{co92,ari96}. The same accuracy as in the 
best variational studies of nuclear matter has been obtained 
in $^{16}$O and $^{40}$Ca  by using spin and isospin 
dependent correlations and modern, microscopic potentials\cite{fab00}. 
To this aim, the nuclear matter single operator chain
approximation\cite{pan79} 
 has been extended to finite nuclear systems to deal with state 
dependent correlations in the framework of the FHNC technique\cite{fab98} 
(FHNC/SOC). The variational and the coupled cluster approaches have provided 
an overall satisfying description of such ground state properties as 
binding energies, one-- and two--body densities and structure functions.  

%
%
In medium heavy nuclei the microscopic approach   
has to be compared with the independent particle
model and its description of various nuclear features. 
The use of this model is so wide and
well--established that great part of the nomenclature on 
medium and heavy nuclei is based upon concepts defined within the IPM
itself. For example, the ideas of collective excitations, single
particle levels, occupation probabilities and spectroscopic
factors are meaningful only in a theoretical framework where the 
IPM is a first order approximation.

The sources of effects beyond the IPM description are generally 
termed as correlations. This name is hiding two rather different kinds
of physical effects. Long range correlations are generated by the
so--called residual interaction, neglected in the IPM and 
reponsible for the collective excitations. The short range 
correlations (SRC) are mainly produced by the repulsive core of the
nucleon--nucleon interaction. Moreover, important non central 
components at intermediate range are needed for most realistic 
descriptions.
While collective phenomena have been known since the early times of
nuclear physics, effects produced by the SRC are more
difficult to be experimentally singled out. 
Only in these last few years, thanks to the advances of the
experimental techniques, there have been some consistent 
effort aimed to their identification.

The sensitivity of the two--nucleon emission cross section to the SRC
is evident \cite{ond97},
but also other quantities related to the behavior of
the single nucleon in the medium seem to depend on them. 
For instance, $(e,e'p)$ data in the quasi-elastic region need a
consistent reduction of the IPM hole strength to be reproduced \cite{dew90}. 
The same holds for the electromagnetic form factors of low lying states 
with high angular momentum \cite{hyd84}.  
Furthermore, charge density distributions obtained by elastic electron
scattering experiments are, in the nuclear interior, smaller than
those predicted by the IPM \cite{cav82}. 
All these facts could be explained by 
assuming occupation probabilities  of the single particle levels different
from that of the IPM \cite{pap86}.
 
From the theoretical point of view, the basic quantity to be
investigated in order to verify the hypothesis of partial occupation
probability is the one body density matrix (OBDM), 
$\rho({\bf r}_1,{\bf r}_{1'})$. 
 defined as:
\begin{equation}
\rho({\bf r}_1,{\bf r}_{1'})=
\langle \Psi_0(A) \vert a^\dagger({\bf r}_1) a({\bf r}_{1'})
       \vert \Psi_0(A) \rangle
\, ,
\label{OBDM_0}
\end{equation}

where $\Psi_0(A)$ is the ground state A--body wave function and 
$a^\dagger({\bf r}_1)$ is the creation operator of a nucleon  
at the position ${\bf r}_1$. 
Much theoretical effort has been devoted to understand
the behavior of the OBDM and of the momentum distribution (MD), 
$n(k)$, given by the Fourier transform of 
$\rho({\bf r}_1,{\bf r}_{1'})$. The MD is also obtained by the 
energy integral of the  spectral function, that 
is often used in plane wave impulse approximation  to study 
inclusive and exclusive reactions. In some approximations 
and kinematics, the MD is directly employed.  
 The natural orbits (NO)\cite{mau91}, with their occupation numbers 
($n_\alpha$),  are defined as  the basis where the OBDM is diagonal. 
In the IPM, the nuclear ground state is described by 
a Slater determinant of fully occupied single particle (SP) wave functions 
below the Fermi surface, $\alpha_F$. In this case, the NO and the SP w.f. 
coincide and $n_{\alpha\leq \alpha_F}$=1 and $n_{\alpha > \alpha_F}$=0. 
 Deviations from this situation are a measure of the correlations, 
since they allow higher NO to become populated with $n_\alpha\neq$0. 

Quantities not directly related to the OBDM, but accessible to
the experimental investigation by means of knock out experiments, are 
the quasihole (QH) wave functions, $\psi_h({\bf r})$,
defined as the overlaps  between $\Psi_0$ and the hole states, 
$\Psi_h$,  produced by removing a nucleon from the position ${\bf r}$.
 From $(e,e'p)$ experiments it is possible to obtain an accurate determination
of the QH overlap functions\cite{kel96}. Their normalizations give the
spectroscopic factors, $S_h$, that deviate from unity
(the IPM value) because of various effects, 
from center of mass to correlation corrections.
Typical values extracted from the experiments are
$S_h\sim 0.6-0.7$\cite{lap93}, and their correct calculation represents
a severe challenge for the available many--body theories.

The FHNC theory for the OBDM in correlated infinite 
matter was developed by Fantoni\cite{fan78}. Subsequentely, FHNC/SOC 
was applied to nuclear matter to calculate both the MD\cite{fan83} 
and the spectral function\cite{ben92}. The technique has been then 
extended to evaluate the OBDM in doubly closed shell nuclei 
described by a correlated wave function containing scalar, 
spin--isospin independent (or Jastrow) correlations\cite{co94}. 
Isospin dependence was later introduced to treat closed shell nuclei 
in the {\em jj} coupling scheme\cite{ari96}. The $^{16}$O momentum 
distribution has been calculated by using state dependent 
correlations\cite{pie92} within the VMC approach. 
Low order cluster expansions have been 
employed in closed shell nuclei\cite{ben86,ari97} 
and in $sp$ and $sd$ shell nuclei with N=Z\cite{mou00}. 
Local density 
approximation was used to estimate the MD in medium--heavy 
nuclei\cite{str90} from the nuclear matter results. The effect 
of short range correlations on the OBDM and the NO in $^{16}$O 
has been studied by means of the Green function method\cite{pol95}. 
As far as the overlap functions are concerned, in the $^7$Li$(e,e'p)^6$He  
reaction the experimental data have been succesfully compared with the VMC, 
parameter--free theoretical predictions\cite{lap99}. VMC has been 
used also in the analysis of the  $^{16}$O$(e,e'p)$ knockout 
experiments\cite{nec98}. A relationship between the OBDM and the 
QH overlaps\cite{nec93} has been recently exploited\cite{gai99,nec97} 
to evaluate the spectroscopic factors for the same reaction, 
using several correlated models for the density matrix. The effect 
of long--range correlations has been considered in Refs.\cite{ami97,gai99}. 

In this paper we extend the FHNC/SOC methodology to deal with the 
OBDM of $^{16}$O and $^{40}$Ca using wave functions with 
central and tensor correlations. The correlated A--body wave function,  
$\Psi_0(1,2...A)$, is written as: 
\begin{equation}
\label{ansatz}
\Psi_0(1,2...A)= G(1,2...A)\Phi_0(1,2...A)
\, , 
\label{Psi_0}
\end{equation}
where  $G(1,2...A)$ is a many--body correlation operator acting on the 
mean field wave function $\Phi_0(1,2...A)$, given by a Slater 
determinant of  single particle wave functions, $\phi_\alpha(i)$. 
The correlation operator is given by a symmetrized product of 
two-body correlation operators, $F_{ij}$, 
\begin{equation}
G(1,2...A)={\cal S}\left[\prod_{i<j}F_{ij}\right]
\, . 
\label{G}
\end{equation}
In the most sophisticated variational calculations (both in nuclear matter 
and in nuclei) $F_{ij}$ assumes the form: 
\begin{equation}
F_{ij}=\sum_{p=1,8}f^p(r_{ij})O^p_{ij}
\, . 
\label{corr8}
\end{equation}
where
\begin{equation}
\label{oper8}
O^{p=1,8}_{ij}=
\left[ 1, \bsigma_i \cdot \bsigma_j, S_{ij}, 
({\bf L} \cdot {\bf S})_{ij} \right]\otimes
\left[ 1, \btau_i \cdot \btau_j \right] 
\end{equation}
and $S_{ij}=(3\,{\hat  {\bf r} }_{ij} \cdot \bsigma_i  \,
{\hat{\bf r}}_{ij} \cdot 
\bsigma_j -  \bsigma_i \cdot \bsigma_j)$ is the tensor operator. 
%
%
The spin--orbit components of the correlation ($p=$7, 8) have been omitted 
in this paper ($f_6$ model), as well as spin--orbit and Coulomb 
interaction terms in the mean field potential generating the 
single particle wave functions, $\phi_\alpha(i)$.  
%
%

The variational principle allows for fixing the correlation functions, 
$f^p(r)$, and the single particle wave functions by minimizing the 
ground state energy,
\begin{equation}
\label{enfun}
E[\Psi_0]=\frac {\langle \Psi_0|H|\Psi_0\rangle}
{\langle \Psi_0|\Psi_0\rangle}
\, . 
\end{equation}
In our calculation we adopt a non relativistic nuclear 
hamiltonian of the form:
\begin{equation}
H={{-\hbar ^2}\over2\,m}\sum_i\nabla_i^2+\sum_{i<j}v_{ij}
+\sum_{i<j<k}v_{ijk} 
\, . 
\label{hamilt}
\end{equation}
Modern two--nucleon potentials, $v_{ij}$, are built on the 
very high quality phase--shift analyses of the 
NN scattering data\cite{arn92,sto93}. We have used the  
 $v'_8$ reduction of the Argonne $v_{18}$ \cite{wir95} potential. 
For the three--nucleon 
interaction the Urbana IX model\cite{pud97} has been adopted.

The plan of the paper is as follows: 
 section 2 is devoted to a brief description of the FHNC/SOC 
theory for the OBDM; in section 3 the results obtained for the 
$^{16}$O and $^{40}$Ca momentum distributions and natural 
orbits are presented and discussed; in section 4 we describe the 
FHNC theory for the overlap functions and give the results 
for the spectroscopic factors; the conclusions are drawn in section 5. 

\section{FHNC/SOC theory for the one body density matrix} 

The one body density matrix (\ref{OBDM_0}) may be written as 
%
\begin{equation}
\rho({\bf r}_1,{\bf r}_{1'})={A\over {\cal N}}
\,\int d^3r_2... \int d^3r_A\,
\Psi^\dagger_0(1,2,..A)\Psi_0(1',2,..A)\, ,
\label{OBDM}
\end{equation}
where ${\cal N}=\int d^3r_1... \int d^3r_A \vert \Psi_0 \vert ^2$. 
The one body density, $\rho_1({\bf r}_1)$,  is 
 the diagonal part of the OBDM, whose  Fourier transform gives 
the momentum distribution,
\begin{equation}
n(k)={1 \over A}\, \int d^3r_1\, \int d^3r_{1'}\,  
\rho({\bf r}_1,{\bf r}_{1'})\, 
e^{i{\bf k} \cdot ({\bf r}_1-{\bf r}_{1'})}\, .
\label{MD}
\end{equation}

In the independent particle model, the OBDM is given by:
\begin{equation}
\rho_{IPM}({\bf r}_1,{\bf r}_{1'})=
\sum_\alpha \phi_\alpha^\dagger(1) \phi_\alpha(1')=
\Bigl (\sum_{\sigma\tau}\chi^\dagger_{\sigma\tau}(1)
\chi_{\sigma\tau}(1')\Bigr )N_0({\bf r}_1,{\bf r}_{1'})\, ,
\label{OBDM_IPM}
\end{equation}
and 
\begin{equation}
\rho_{1,IPM}({\bf r}_1)=
\sum_\alpha \vert \phi_\alpha(1)\vert ^2=
\nu N_0({\bf r}_1,{\bf r}_1)\, ,
\label{OBD_IPM}
\end{equation}
where $\chi_{\sigma\tau}(1)$ is the spin--isospin single particle 
eigenfunction and $\nu$ the degeneracy number ($\nu=4$, for the 
doubly closed shell, N=Z nuclei we are considering). The second 
equalities in the above equations are valid since we are working in 
$ls$ coupling.  

In analogy, we define the function $N({\bf r}_1,{\bf r}_{1'})$ via the relation:
\begin{equation}
\rho({\bf r}_1,{\bf r}_{1'})=
\Bigl (\sum_{\sigma\tau}\chi^\dagger_{\sigma\tau}(1)
\chi_{\sigma\tau}(1')\Bigr )N({\bf r}_1,{\bf r}_{1'})\, .
\label{OBDM_1}
\end{equation}
We have already presented in Ref.\cite{co94}
the FHNC theory of the OBDM for Jastrow correlated wave functions,  
\begin{equation}
G_J(1,2...A)=\prod_{i<j}f^1(r_{ij})\, ,
\label{G_J}
\end{equation}
having only the $p=1$ scalar component in the correlation factor $F_{ij}$.    
The density matrix is expanded in powers 
of the {\sl dynamical correlations}, $h(r)=[f^1(r)]^2-1$ ($h$--bond) 
and $\omega(r)=f^1(r)-1$ ($\omega$--bond), and of the {\sl statistical 
correlations}, $N_0({\bf r}_i,{\bf r}_j)$ (exchange or $e$--bonds). 
The expansion generates  cluster terms classified according to the 
number of particles and to the number of the correlations. 
The FHNC equations allow for summing cluster terms at all orders. 
Details of the finite systems FHNC theory may be found in \cite{co94} and 
in Ref.\cite{co92}. 

The FHNC/SOC equations for the more general $f_6$ correlation 
were derived in Ref.\cite{fab98} for the one and two 
body densities. In the case of the OBDM, we can write  
 \begin{eqnarray}
 N({\bf r}_1,{\bf r}_{1'})&=&
  \xi^c_\omega({\bf r}_1)[1+\Delta\xi^{op}_\omega({\bf r}_1)]
  \xi^c_\omega({\bf r}_{1'})[1+\Delta\xi^{op}_\omega({\bf r}_{1'})]
 e^{N_{\omega\omega}^c({\bf r}_1,{\bf r}_{1'})} 
 \\ \nonumber
 &\times&\Bigl [N_0({\bf r}_1,{\bf r}_{1'})
 -N_{\omega_c \omega_c}^c({\bf r}_1,{\bf r}_{1'})\Bigr ]
 +\xi^c_\omega({\bf r}_1)\xi^c_\omega({\bf r}_{1'})
 e^{N_{\omega\omega}^c({\bf r}_1,{\bf r}_{1'})} 
 \\ \nonumber
 &\times&\sum_{p\geq 2}A^p\Delta^p
 \Bigl \{ N_{\omega\omega}^p({\bf r}_1,{\bf r}_{1'}) 
 \Bigl [ N_0({\bf r}_1,{\bf r}_{1'})
 -N_{\omega_c \omega_c}^c({\bf r}_1,{\bf r}_{1'})\Bigr ]
 -N_{\omega_c \omega_c}^p({\bf r}_1,{\bf r}_{1'})\Bigr \}
 \, .
 \label{N_SOC}
 \end{eqnarray}
$N_{\omega\omega}^p$ and $N_{\omega_c \omega_c}^p$ are {\sl nodal} 
functions (see \cite{co94}) 
of the ${\omega\omega}$ and ${\omega_c\omega_c}$ 
type, $\xi^c_\omega$ and $\Delta\xi^{op}_\omega$ are the central 
vertex corrections and their operatorial contributions\cite{fan83} and 
the matrices $A^p$ and $\Delta^p$ are defined in \cite{fab98}. 
The components will be often labelled as $c$ ($p=1$) and $\sigma$ 
(spin), $\tau$ (isospin) and $t$ (tensor).

The nodal funtions are obtained by the equations:
\begin{equation}
N^p_{\omega\omega}({\bf r}_1,{\bf r}_{1'})
=\sum_{xx'}\sum_{qr}
\int d^3r_2  \xi^{qrp}_{121'} X^q_{\omega x}({\bf r}_1,{\bf r}_2)
V^{qr}_{xx'}({\bf r}_2)
\left[ X^r_{x'\omega}({\bf r}_2,{\bf r}_{1'})
     + N^r_{x'\omega}({\bf r}_2,{\bf r}_{1'})
\right]  
 \, ,
\label{Nww_SOC}
\end{equation}
and
 \begin{eqnarray}
N^p_{\omega_c\omega_c}({\bf r}_1,{\bf r}_{1'})
&=&\sum_{qr}
\int d^3r_2  \xi^{qrp}_{121'} X^q_{\omega_c c}({\bf r}_1,{\bf r}_2)
V^{qr}_{cc}({\bf r}_2)
\left[ X^c_{c\omega_c}({\bf r}_2,{\bf r}_{1'})
     + N^c_{c\omega_c}({\bf r}_2,{\bf r}_{1'})
\right] \Delta^r  
 \\ \nonumber
&+&\sum_{qr\geq 2}
\int d^3r_2  \xi^{qrp}_{121'} \Delta^q X^c_{\omega_c c}({\bf r}_1,{\bf r}_2)
V^{qr}_{cc}({\bf r}_2)
\left[ X^r_{c\omega_c}({\bf r}_2,{\bf r}_{1'})
     + N^r_{c\omega_c}({\bf r}_2,{\bf r}_{1'})
\right] 
 \\ \nonumber
&+&N^p_{\rho\rho}({\bf r}_1,{\bf r}_{1'})
 + N^p_{\rho\omega_c}({\bf r}_1,{\bf r}_{1'})
 + N^p_{\omega_c\rho}({\bf r}_1,{\bf r}_{1'})
 \, .
\label{Nwcwc_SOC}
\end{eqnarray}
The index $x(x')$ may assume the values $x(x')=d,e$ and denotes the 
exchange pattern and the type of correlation at a specific point. 
In general, we may have: $d$-- and $\omega$-vertices, if the 
point is reached by $h$-- and $\omega$-bonds, 
respectively, and does not belong to any exchange loop;
an $e$--vertex if the point belongs to a closed exchange loop 
and it is reached by two $e$--bonds;  
$c$-- and $\omega_c$--vertices if the point belongs to an open 
exchange loop and it is reached by a single exchange line. 
The allowed $(xx')$ combinations 
are: $dd$, $de$, $ed$. We indicate with $V^{qr}_{xx'}$ the vertex corrections, 
and with $\xi^{qrp}_{121'}$ the angular couplings. 
The expressions of these equations are all given in \cite{fab98}. 

The above equations are derived in the FHNC/0 approximation, 
which does not include the contribution of the {\sl elementary} 
diagrams. A detailed discussion of the importance of these diagrams and 
of the accuracy of the approximations used in solving the FHNC equations 
can be found in Refs.\cite{fab92,viv86}. The FHNC/0 choice 
has been used in studies of the equation of state of nuclear  
matter\cite{wir88}, where it was found that the elementary diagrams 
contribution is not important because of the relatively low densities
of the system. In finite nuclei elementary diagrams may play  
some role in the evaluation of the expectation value of potentials 
having strong exchange components\cite{co92}. In general, a measure of 
the relevance of the missing diagrams is provided by the accuracy of 
the sum rules of the one and two body densities.

The partial nodal functions, $N^p_{\omega x}$, are solutions of 
the integral equations: 
\begin{equation}
N^p_{\omega x}({\bf r}_1,{\bf r}_2)=\sum_{yy'}\sum_{qr}
\int d^3r_3  \xi^{qrp}_{132} X^q_{\omega y}({\bf r}_1,{\bf r}_3) 
V^{qr}_{yy'}({\bf r}_3)
 \left[ X^r_{y'x}({\bf r}_3,{\bf r}_2) 
+ N^r_{y'x}({\bf r}_3,{\bf r}_2) \right]  
 \, .
\label{Nwx_SOC}
\end{equation}
The $X^{p\geq 2}_{\omega x}$ links are
\begin{equation}
X^{p\geq 2}_{\omega d}({\bf r}_1,{\bf r}_2)=
h^p_\omega ({\bf r}_1,{\bf r}_2)
h^c_\omega({\bf r}_1,{\bf r}_2) - 
N^p_{\omega d}({\bf r}_1,{\bf r}_2) 
\, ,
\label{Xwd}
\end{equation}
\begin{equation}
X^p_{\omega e}({\bf r}_1,{\bf r}_2)=
h^c_\omega({\bf r}_1,{\bf r}_2) 
\left [ h^p_\omega ({\bf r}_1,{\bf r}_2) 
N^c_{\omega e}({\bf r}_1,{\bf r}_2) + 
f^c(r_{12}) 
N^p_{\omega e}({\bf r}_1,{\bf r}_2) \right ] - 
N^p_{\omega e}({\bf r}_1,{\bf r}_2) 
\, ,
\label{Xwe}
\end{equation}
\begin{equation}
 h^p_\omega({\bf r}_1,{\bf r}_2)=
  f^p(r_{12}) + f^c(r_{12}) N^p_{\omega d}({\bf r}_1,{\bf r}_2) 
\, ,
\label{hp}
\end{equation}
and $h^c_\omega({\bf r}_1,{\bf r}_2) =\exp  
\left[ N^c_{\omega d}({\bf r}_1,{\bf r}_2)  \right]$. Moreover, the 
$ N(X)^p_{xy}({\bf r}_1,{\bf r}_2)=N(X)^p_{yx}({\bf r}_2,{\bf r}_1)$ 
property holds. The central $X^{c}_{\omega x}$ links are defined in
\cite{co94},  
and the $X^{p}_{xy}$ and $N^{p}_{xy}$ functions in \cite{fab98}. 

For the $\omega_cc$--type nodals we have
$N^p_{\omega_cc}=N^p_{\omega_cx}+N^p_{\omega_c \rho}+
N^p_{\rho x}+N^p_{\rho\rho}$. $N^p_{\omega_cx}$ 
and $N^p_{\omega_c \rho}$ are solutions of
\begin{eqnarray}
N^p_{\omega_cx}({\bf r}_1,{\bf r}_2) &=&
\sum_{qr}
\int d^3r_3  \xi^{qrp}_{132} X^q_{\omega_cc}({\bf r}_1,{\bf r}_3) 
V^{qr}_{cc}({\bf r}_3) 
 \left[ X^c_{cc}({\bf r}_3,{\bf r}_2) 
 + N^c_{xx}({\bf r}_3,{\bf r}_2)  
 + N^c_{\rho x}({\bf r}_3,{\bf r}_2) 
\right] \Delta ^r    \\ \nonumber &+&   
 \sum_{qr\geq 2}
\int d^3r_3  \xi^{qrp}_{132} \Delta ^q 
 X^c_{\omega_cc}({\bf r}_1,{\bf r}_3) 
 V^{qr}_{cc}({\bf r}_3)
 \left[ X^r_{cc}({\bf r}_3,{\bf r}_2) 
 + N^r_{xx}({\bf r}_3,{\bf r}_2) 
 + N^r_{\rho x}({\bf r}_3,{\bf r}_2) \right]  
\, ,
\label{Nwcc_wx}
\end{eqnarray}
\begin{eqnarray}
N^p_{\omega_c \rho }({\bf r}_1,{\bf r}_2) &=&
\sum_{qr}
\int d^3r_3  \xi^{qrp}_{132} X^q_{\omega_cc}({\bf r}_1,{\bf r}_3) 
 V^{qr}_{cc}({\bf r}_3)
 \left[ -N_0({\bf r}_3,{\bf r}_2)  
 + N^c_{x\rho }({\bf r}_3,{\bf r}_2) 
 + N^c_{\rho\rho}({\bf r}_3,{\bf r}_2) \right] \Delta ^r    
 \\ \nonumber &+&   
 \sum_{qr\geq 2}
 \int d^3r_3  \xi^{qrp}_{132} \Delta ^q 
 X^c_{\omega_cc}({\bf r}_1,{\bf r}_3) 
 V^{qr}_{cc}({\bf r}_3)
 \left[ N^r_{x\rho }({\bf r}_3,{\bf r}_2) 
 + N^r_{\rho\rho  }({\bf r}_3,{\bf r}_2) \right]  
\, .
\label{Nwcc_wp}
\end{eqnarray}
The equations for $N^p_{\rho x}$ and $N^p_{\rho\rho}$ are given in
\cite{fab98}. 
The links $X^{p\geq 2}_{\omega_cc}$ are
\begin{eqnarray}
X^{p\geq 2}_{\omega_cc}({\bf r}_1,{\bf r}_2)
&=&h^p_\omega({\bf r}_1,{\bf r}_2)h^c_\omega ({\bf r}_1,{\bf r}_2) 
 \left[ N^c_{\omega_cc}({\bf r}_1,{\bf r}_2) 
- N_0({\bf r}_1,{\bf r}_2) \right] 
 \\ \nonumber &+&   
 \left [ f^c(r_{12}) h^c_\omega ({\bf r}_1,{\bf r}_2) -1 \right ] 
 N^p_{\omega_cc}({\bf r}_1,{\bf r}_2) 
\, .
\label{Xwcc}
\end{eqnarray}
Again, $X^c_{\omega_cc}$ is defined in \cite{co94}.  
 
 The vertex corrections, $\xi^c_\omega$, are discussed in \cite{co94} and 
$\Delta\xi^{op}_\omega$ is given by:
\begin{eqnarray}
\Delta\xi^{op}_\omega({\bf r}_1)&=&
U^{op}_\omega({\bf r}_1)+ \sum_{p\geq 2}{A^p \over 2}
\int d^3r_2  
 \left \{ 
 X^p_{\omega d}({\bf r}_1,{\bf r}_2)
 N^p_{\omega d}({\bf r}_1,{\bf r}_2) 
 \rho_1^{c}({\bf r}_2)\right . 
 \\ \nonumber &+&   
 \left . \left[ 
 X^p_{\omega d}({\bf r}_1,{\bf r}_2)
 N^p_{\omega e}({\bf r}_1,{\bf r}_2) + 
 X^p_{\omega e}({\bf r}_1,{\bf r}_2)
 N^p_{\omega d}({\bf r}_1,{\bf r}_2) \right ]
 C_d({\bf r}_2)\right\} 
 \, .
\label{Delta_csi}
\end{eqnarray}
$\rho_1^{c}$ (the Jastrow part of the one body density) and 
$C_d$ are given in \cite{fab98}, and $U^{op}_\omega$ is obtained by 
eq.(2.12) of \cite{co94} with the substitutions 
$\rho_1^{c}\rightarrow \rho_1 -\rho_1^{c}$ and 
$C_d\rightarrow C_d U_d^{op}$. 
\section{Results for the momentum distribution and natural orbits}

In our work we used the Argonne  $v'_{8}$ NN potential. This model 
is  based upon the $v_{18}$ potential and it is constructed by 
considering only the first eight operator terms, up to the spin--orbit 
ones. It reproduces the isoscalar part of the full interaction, $v_{18}$,  
in the $S$, $P$ and $^3D_1$ waves and the $^3D_1$--$^3S_1$ coupling.
Argonne  $v'_{8}$ was introduced in Ref.\cite{pud97} 
because its parametrization (simpler than that of other realistic 
potentials, since $L^2$ and $(L\cdot S)^2$ components are missing) 
allowed for a large simplification in the numerically involved quantum Monte 
Carlo calculations.  
The $v'_{8}$ potential is slightly more attractive than $v_{18}$, and, 
for this reason, the strength of the associated repulsive part of the 
three--nucleon force (Urbana IX model) has been increased by 30$\%$ 
with respect to the original version. 
The results presented in this paper have been obtained with this 
hamiltonian (A8'+UIX model). 

The correlation functions, $f^p(r)$, and the single particle functions, 
$\phi_\alpha(i)$, are the two ingredients necessary to construct
the many--body wave function (\ref{Psi_0}). We use  
a $f_6$ correlation, therefore, with respect to the structure of the
hamiltonian, we neglect the spin--orbit components. 
The correlation is determined by minimizing the  nuclear matter
energy at the lowest order of the cluster expansion,
considering the Fermi momentum, $k_F$, as one of the variational parameters.
The resulting two--body Euler equations are solved with 
the {\em healing} conditions $f^{1}(r\geq d_{1})=1$, 
$f^{p>1}(r\geq d_{p})=0$, and 
requiring that the first derivatives vanish at $r=d_p$. 
Only two healing distances are introduced, $d_c$ for the four central 
channels and $d_t$ for the tensor ones, and they are 
variationally fixed. More details on this procedure are given in 
Ref.\cite{pan79} for nuclear matter and in \cite{fab98} for nuclei. 

The single particle wave functions have been obtained by solving 
the single particle Schr\"odinger equation with a Woods--Saxon potential, 
\begin{equation}
V_{WS}(r)= {V_0 \over {1+\exp{[(r-R_0)/a_0]}}}\, .
\label{WS}
\end{equation}

A full minimization for the A8'+UIX model has been obtained in
Ref. \cite{fab00} and it has provided a binding energy per 
nucleon, B/A,  of 5.48 MeV in $^{16}$O and 6.97 MeV in $^{40}$Ca, 
to be compared with the experimental values of 7.97 MeV ($^{16}$O) 
and 8.55 MeV ($^{40}$Ca). 
These differences are comparable with those obtained in nuclear 
matter at the empirical saturation density, $\rho_{NM}=$0.16 fm$^{-3}$, 
with the same hamiltonian. In fact, the  FHNC/SOC nuclear matter 
energy per nucleon, $E_{NM}$, is $E_{NM}=$-10.9 MeV\cite{fab00},  
against the empirical value of -16 MeV. 

The nuclear root 
mean square radii (rms) were found to be 2.83 fm in $^{16}$O and 3.66 fm 
in $^{40}$Ca (the experimental radii are 2.73 fm and 3.48 fm, respectively).
However, the one body densities at the variational minimum did not show 
a satisfactory agreement with the experimental ones. Moreover, shallow 
minima with respect to variations of the mean field parameters 
around the minimum itself were found in Ref.\cite{fab00}. In particular, 
if one chooses a set of single particle wave functions which reproduces at 
best the empirical densities, the A8'+UIX model provided 
B/A=5.41 MeV in $^{16}$O and B/A=6.64 MeV in $^{40}$Ca, with 
rms($^{16}$O)=2.67 fm and rms($^{40}$Ca)=3.39 fm. Therefore, the density 
description has largely improved while the energy variations are kept 
within the accuracy of the FHNC/SOC scheme. The results presented 
in this paper have been obtained by means of this type of wave 
function, whose parameters are given in Table V of Ref.\cite{fab00}.

\begin{figure}
\epsfig{file=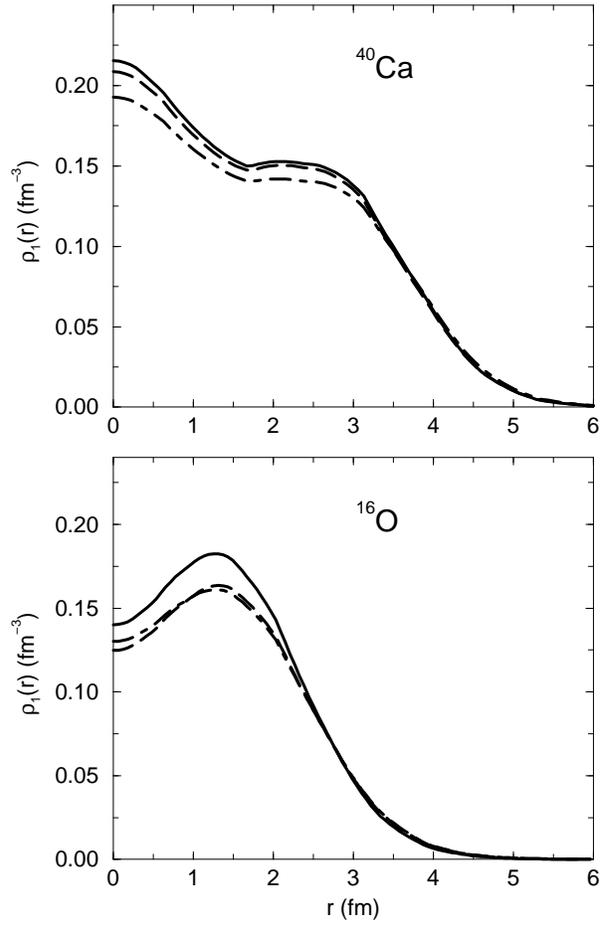,width=15 cm}
\caption{
Nuclear one body densities. The solid lines are the FHNC/SOC 
results with the $f_6$ correlation, the dot--dashed lines are the 
densities with the Jastrow correlation, the dashed lines 
are the IPM densities.
}
\label{fig:fig1}
\end{figure}

The one body densities generated by the FHNC/SOC scheme are shown 
in Fig. \ref{fig:fig1}, where the solid lines give the densities 
obtained with the full correlation, the dot--dashed lines are those obtained 
with the Jastrow correlation (retaining only the $p=$1 component) and 
the dashed lines are the IPM densities. The effect of the operatorial 
correlation is large with respect to the Jastrow case, that is hardly 
distinguishable from the IPM one. The comparison with the experimental 
results has been presented in Ref.\cite{fab00}, where the proton densities 
are folded with the electromagnetic nucleon form factor.

\begin{figure}
\epsfig{file=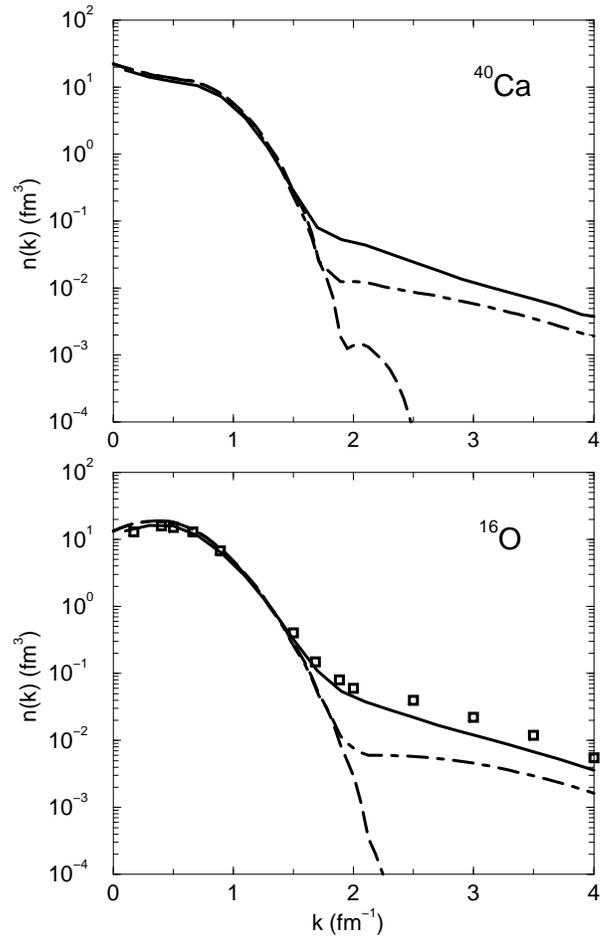,width=15 cm}
\caption{
FHNC/SOC momentum distributions in $^{16}$O and $^{40}$Ca.  
Solid lines: $f_6$ model; dot--dashed: Jastrow; 
dashed: IPM. The squares are the VMC results of Ref.\protect\cite{pie92}.
}
\label{fig:fig2}
\end{figure}

The momentum distributions are given in Fig. \ref{fig:fig2}. Again the 
solid and dot--dashed lines are the fully correlated and Jastrow results, 
respectively, while the IPM ones are shown as dashed lines. The 
squares are the VMC results\cite{pie92} for $^{16}$O obtained with 
the Argonne $v_{14}$\cite{wir84} NN interaction. 

The MD is normalized as:
\begin{equation}
 1\,=\,{\nu\over{(2\pi)^3}}\int d^3 k \, n(k)\, ,
\label{norm}
\end{equation}
where $\nu$ is the spin--isospin degeneracy.
For the Jastrow correlation this 
normalization is satisfied within the  0.2--0.3 $\%$ 
in both nuclei, while for the $f_6$ model the error is 
$\sim$3 $\%$ in $^{16}$O and $\sim$2 $\%$ in $^{40}$Ca, reflecting 
the approximations of the SOC approach.

\begin{figure}
\epsfig{file=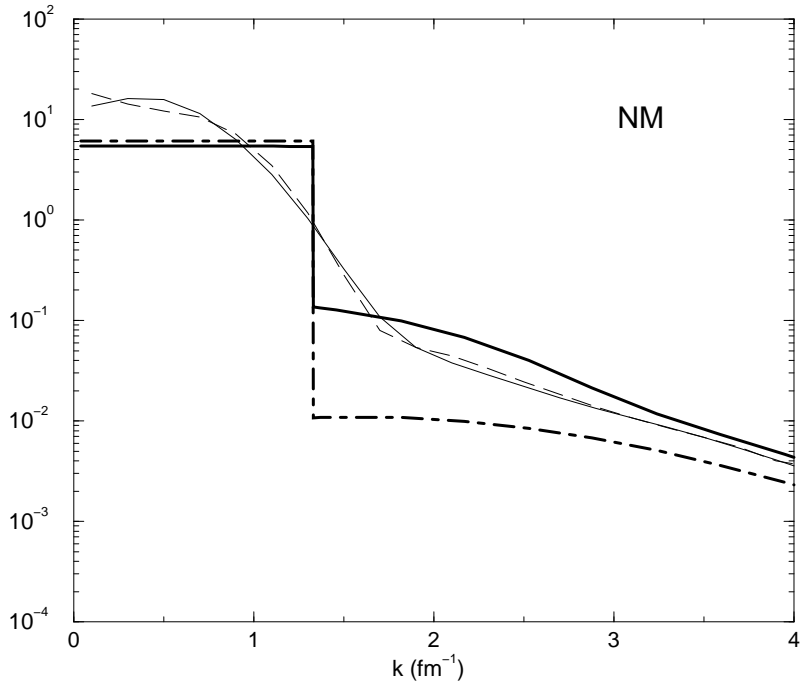,width=15 cm}
\caption{
FHNC/SOC momentum distributions in $^{16}$O, $^{40}$Ca and nuclear 
matter (NM). 
Solid line: NM $f_6$ model; dot--dashed: NM Jastrow model; 
thin solid line: $^{16}$O $f_6$ model; 
thin dashed line: $^{40}$Ca $f_6$ model. 
}
\label{fig:fig2a}
\end{figure}

In Fig. \ref{fig:fig2a}  the MD of $^{16}$O (thin continuous
line) and that of  $^{40}$Ca (thin dashed line) are compared with
those of nuclear matter normalized as in (\ref{norm}) 
and calculated  in FHNC/SOC framework by using the same interaction. 
It is worth noticing that the differences between the Jastrow and the 
$f_6$ correlations are similar in the infinite and finite 
systems and that the three cases show an analogous behavior at 
large momentum values. 
This momentum region is dominated by the short range structure of the 
nuclear wave function, which is heavily affected by the NN correlations. 
The effect appears to be, to a large extent, 
independent on the nucleus. A similar behavior was found in Ref.\cite{pie92}, 
where the comparison was made among the $^{4}$He, $^{16}$O and nuclear 
matter momentum distributions. With respect to the Jastrow estimates
the non central, tensor correlations enhance 
the tails of the MDs by a factor 2--3 slightly smaller than the one
found in Ref.\cite{pie92}, which is roughly $\sim$4. 
The difference may be understood in terms of the stronger 
tensor force of the Argonne $v_{14}$ potential adopted in that reference. 
%
%
Part of the discrepancy may also be ascribed to the presence of 
spin--orbit correlations in the wave function of Ref.\cite{pie92}. However, 
we notice that, in the same paper, it was found that these correlations 
contribute to the kinetic energy by only $\sim 1\%$ (0.4 MeV/A out of 
a total kinetic energy of 34.4 MeV/A).
%
%

A more demanding sum rule for $n(k)$ than that expressed by 
eq.(\ref{norm}) can be obtained from the kinetic energy, $T$.
The kinetic energy per particle can be evaluated via the MD as:
\begin{equation}
{T\over A}= {{\hbar^2}\over {2m}} {\nu\over{(2\pi)^3}}\int d^3 k \,k^2 n(k)
\equiv T_{MD}
\, .
\label{kinet}
\end{equation}
The value of
$T/A$ can also be computed in the FHNC/SOC framework ($T_{FHNC}$), for
example by means of the Jackson-Feenberg identity, as it has been done in
Ref.\cite{fab00} for the A8'+UIX model. 
The differences between $T_{MD}$ and $T_{FHNC}$ are a severe 
measure of the importance of the approximations made in the 
cluster expansion (FHNC/0 and SOC). 
For the Jastrow cases the relative disagreement, 
$\delta T= \vert T_{MD}-T_{FHNC}\vert /T_{FHNC}$, is $<5\%$ (
$T_{MD}$=20.52 MeV and $T_{FHNC}$=19.57 MeV in $^{16}$O and 
$T_{MD}$=22.98 MeV and $T_{FHNC}$=22.05 MeV in $^{40}$Ca) 
%
%
 and it is due to the absence of the elementary diagrams in the FHNC/0 
 truncation.
%
%
In the  $f_6$ model we obtained 
$T_{MD}$=29.42 MeV, $T_{FHNC}$=32.64 MeV and 
$\delta T=$9$\%$ 
in $^{16}$O and $T_{MD}$=36.63 MeV, $T_{FHNC}$=38.15 MeV and 
$\delta T=$9.6$\%$ in $^{40}$Ca. 
This larger disagreement is due to the SOC approximation. 
The $T_{MD}$ value is largely influenced by the 
behavior of the momentum distribution at high $k$--values.  
The contribution of the $k>$5 fm$^{-1}$ tail has been evaluated 
by an exponential extrapolation of the computed MD. 
%
%
The tail contributions for the Jastrow and operatorial  correlations 
are about 5$\%$ and 10$\%$ of the total $T_{MD}$, respectively. 
So, we believe that the uncertainty in $T_{MD}$ related to the MD tails 
may be fixed to a few percent in both cases.
As a additional check of the numerical accuracy of the algorithm used to 
evaluate the momentum distribution, 
%
%
we have verified that $T_{MD}$ coincides with $T_{FHNC}$ for the IPM.

The NO and their occupation numbers
are obtained by diagonalizing the OBDM:
\begin{equation}
\rho_1({\bf r}_1,{\bf r}_{1'})=\sum_\alpha
 n_\alpha
 \phi_\alpha^{NO}({\bf r}_1)^\dagger
 \phi_\alpha^{NO}({\bf r}_{1'})
\, ,
\label{natural}
\end{equation}

We treat spherical nuclei in $ls$ single particle coupling, saturated
in both spin and isospin. For this reason the spin--isospin part of
eq. (\ref{OBDM_1}) provides the degeneracy $\nu$=4. Because of the
spherical symmetry the function $N({\bf r}_1,{\bf r}_{1'})$ of 
eq. (\ref{OBDM_1}) can be expanded in multipoles, and we obtain
for the OBDM:
\begin{equation}
\rho_1({\bf r}_1,{\bf r}_{1'})= \nu
\sum_l {{2l+1}\over {4\pi}}
P_l(\cos \theta_{11'}) \rho_l(r_1,r_{1'})
\, ,
\label{natural_1}
\end{equation}
where $P_l(x)$ represents the Legendre polinomials and $\theta_{11'}$
is the angle between ${\bf r}_1$ and ${\bf r}_{1'}$.

Exploiting again the spherical symmetry, the natural orbitals 
can be written as: 
\begin{equation}
 \phi_{\alpha=nlm}^{NO}({\bf r})
  = \phi_{nl}^{NO}(r) \, Y_{lm}({\hat r}) \chi_{\sigma \tau}
\, ,
\label{natural_2}
\end{equation}
where we indicate with $Y_{lm}({\hat r})$ the spherical harmonics and 
with $\chi_{\sigma \tau}$ the spin--isospin part of the wave function. 
The normalization condition is:
\begin{equation}
1 \, = \, \int r^2 \, dr \vert \phi_{nl}^{NO}( r)\vert ^2
\, .
\label{natural_4}
\end{equation}
Therefore we obtain:
\begin{equation}
\rho_l(r_1,r_{1'})= \nu \sum_n n_{nl}\,
  \phi_{nl}^{NO}( r_1) \, \phi_{nl}^{NO}( r_{1'})
\, .
\label{natural_3}
\end{equation}
The $nl$--natural orbitals and their occupations  have been 
obtained by 
%
%
discretizing and diagonalizing the matrix $\rho_l(r_1,r_{1'})$ 
in a $100\times 100$ equally spaced grid, up to $r_{max}=6(7)$ fm for 
$^{16}$O ($^{40}$Ca).  
%
%

\begin{figure}
\epsfig{file=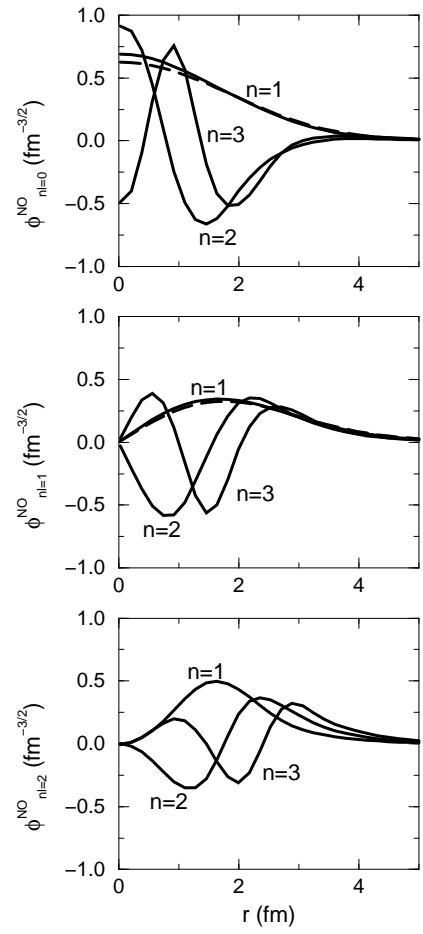,width=15 cm}
\caption{
$^{16}$O natural orbits.
Solid lines: $f_6$ model; dashed: IPM. 
}
\label{fig:fig3}
\end{figure}

\begin{figure}
\epsfig{file=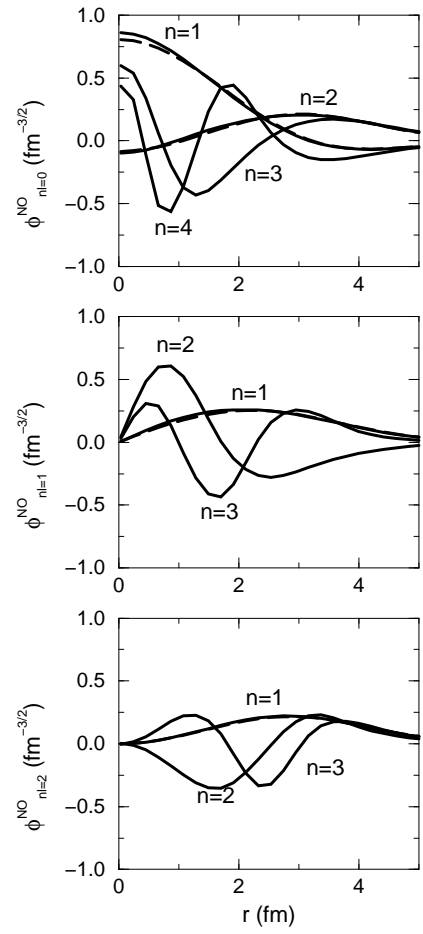,width=15 cm}
\caption{
$^{40}$Ca natural orbits.
Solid lines: $f_6$ model; dashed: IPM. 
}
\label{fig:fig4}
\end{figure}

The first three NO of $^{16}$O and $^{40}$Ca, calculated for the
three lowest $l$--values  are shown in Figs. \ref{fig:fig3} and
\ref{fig:fig4}. 
In Table \ref{tab:tab1} 
the occupation numbers of the various NO for Jastrow and $f^6$
correlations are presented. 

The effect of 
the correlations on the shell model orbitals are mainly visible in 
the short range part of the $1s$ state, making this NO more localized 
than its IPM counterpart. The shape of the other IPM states is 
barely influenced by the correlations. The occupation of the 
NO corresponding to the shell model ones is depleted 
by as much as 22$\%$ (the $2s$ state in $^{40}$Ca). In contrast,  
the mean field unoccupied  states become sizeably populated. 
The two effects are largely due to the tensor part of the correlation 
operator.

It was pointed out in Ref.\cite{lew88}, that when there is more than 
one occupied state in the IPM for a given $l$--value (as for the 
$s$--states in $^{40}$Ca), then the natural orbitals may be 
qualitatively different from the IPM ones. In fact, 
any orthogonal combination of the mean field 
orbitals does not change the corresponding density matrix. New 
mean field $s$--orbitals, $\hat\phi_{ns}$, 
consistent with  $\rho_{IPM}({\bf r}_1,{\bf r}_{1'})$, 
can be obtained in $^{40}$Ca by the transformations: 
\begin{equation}
\hat\phi_{1s}({\bf r})=\cos(\alpha)\phi_{1s}({\bf r})
                      +\sin(\alpha)\phi_{2s}({\bf r})
\, ,
\label{loc1}
\end{equation}
\begin{equation}
\hat\phi_{2s}({\bf r})=\cos(\alpha)\phi_{2s}({\bf r})
                      -\sin(\alpha)\phi_{1s}({\bf r})
\, ,
\label{loc2}
\end{equation}
for any choice of the angle $\alpha$. The IPM orbitals of the $l$=0 
panel in Fig. \ref{fig:fig4} are obtained by a numerical diagonalization 
of the one--body density matrix and roughly correspond to 
$\cos(\alpha)=.8$.

The $^{16}$O natural orbits have been evaluated in Ref.\cite{pol95} 
within a Green function approach and using the one--boson--exchange 
Bonn B potential  of Ref.\cite{mac89}. The authors find the $n=$1 NO 
more populated than the CBF ones for the 
occupied states in the shell model approach ($l=$0, 1), and, 
consequently, lower occupations for all the remaining orbitals. 
Specifically, the $n$=1 Green function results are: 
$n_{1s}$=.921,  $n_{1p}$=.941 and  $n_{1d}$=.017. The 
$1p$ ($1d$) occupation number has been taken as the 
average of the $1p_{1/2}$ and $1p_{3/2}$ 
($1d_{3/2}$ and $1d_{5/2}$) orbitals given in the reference. 
The discrepancies are probably to be ascribed more to the different 
potentials adopted, rather than to the methodologies. The 
A8'+UIX model induces stronger correlation, so giving a 
larger depletion of the lowest NO. This effect was also found 
in the study of $^3$He atomic drops of Ref.\cite{lew88}, where 
the strong repulsive interaction between the $^3$He atoms 
depletes the shell model occupations by 15--46 $\%$. 
The CBF total occupation numbers in the $l$--th 
orbitals for $^{16}$O and $^{40}$Ca with different 
correlations ($f_6$, $f_4$, without tensor components, and 
Jastrow, J) are given in Table \ref{tab:tab2}, 
together with the $^{16}$O Green function ones from Ref.\cite{pol95}. 
It appears clearly that the longer ranged tensor correlations are 
responsible of most of the deoccupation of the shell model natural 
orbitals in favour of the higher ones.

\begin{figure}
\epsfig{file=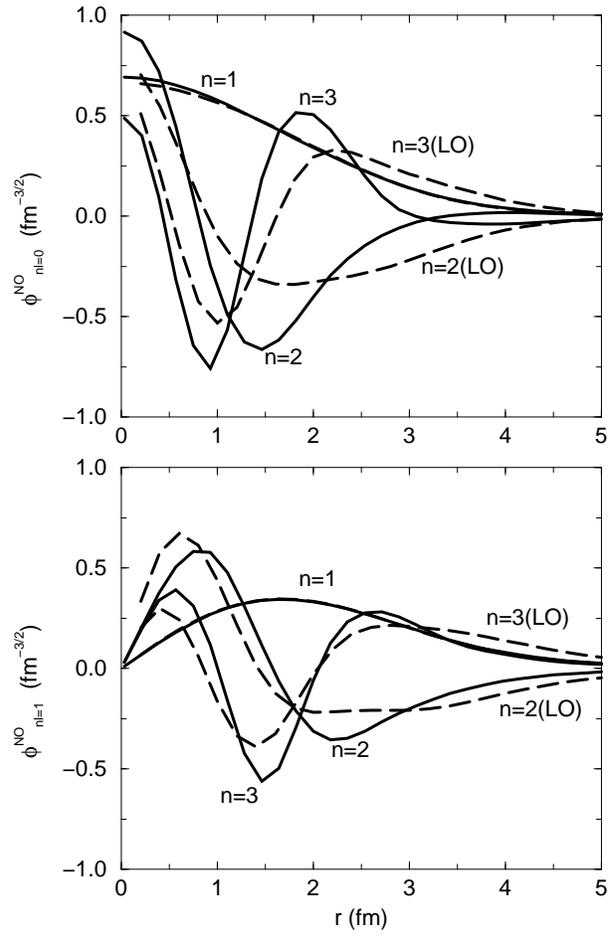,width=15 cm}
\caption{
$^{16}$O $l=$0, 1 natural orbits in FHNC/SOC (solid lines) 
and LO approximation  (dashed lines).
}
\label{fig:fig5}
\end{figure}

Fig. \ref{fig:fig5} presents a comparison between the $^{16}$O NO 
in FHNC/SOC and in the lowest order approximation. This approach 
consists in truncating the cluster expansion at the first order in the 
dynamical correlation lines\cite{stoi93,van97,ari97}
and it has achieved a certain degree of popularity 
because of its simplicity. 
The approximation provides a good description 
of the $n$=1 NO corresponding to the occupied shell model states, 
 but it fails to reproduce the other ones. 

\begin{figure}
\epsfig{file=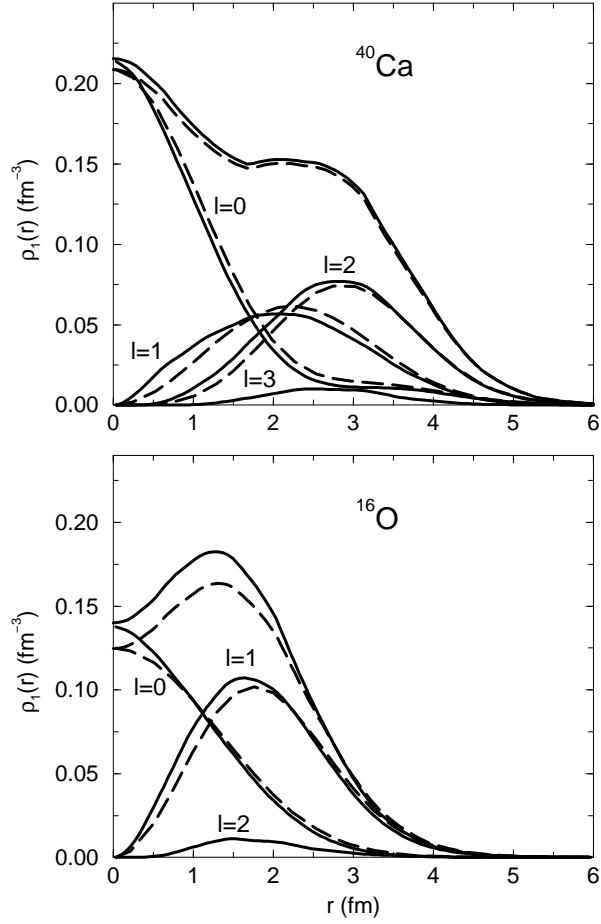,width=15 cm}
\caption{
Partial wave contributions to the one body densities in the natural 
orbits representation. The upper lines are the total  densities. 
The remaining lines give $\nu \frac{ 2l+1}{4\pi} \rho_l(r,r)$. 
Solid lines: $f_6$ model; dashed: IPM. 
}
\label{fig:fig6}
\end{figure}

To conclude this section, we give in Fig. \ref{fig:fig6} the partial wave 
decomposition of the correlated and IPM one body densities in terms of
their natural orbits.

\section{Quasi hole states and single particle overlap functions }

A considerable amount of information on the properties of the 
single nucleon in the nuclear medium can be deduced from 
$(e,e'p)$ reactions. These experiments have been analyzed to extract
the quasihole function, $\psi_h({\bf r})$, given by the 
overlap  between the A--body ground--state and the (A-1)--body 
hole state of the residual system.

In a fixed center reference frame, the QH function is defined as:
\begin{equation}
\psi_h(x) = {\sqrt A} {
{\langle \Psi_h(A-1)\vert \delta (x-x_A) \vert \Psi_0(A)\rangle}
\over
{\langle \Psi_h(A-1)\vert \Psi_h(A-1)\rangle ^{1/2}
 \langle \Psi_0(A)\vert \Psi_0(A)\rangle ^{1/2}}
}
\, ,
\label{QH}
\end{equation}
In doubly closed shell nuclei in the $ls$ coupling scheme
it is possible to separate the radial dependence of the QH function
from the angular, spin and isospin ones, as:
\begin{equation}
\psi_h(x) = \psi_h(r) Y_{lm}({\hat r})\chi_{\sigma \tau}
= \psi_h(r) 
{\cal Y}_{lm\sigma \tau}({\hat r})
\, .
\label{psi_h}
\end{equation}
In the IPM, the QH overlaps are simply the shell 
model functions and $\psi^{IPM}_h(r) = R_{h=nl}(r)$, where
$R_{h}(r_i)$ is the radial part of $\phi_\alpha(i)$. 

In CBF theory $\Psi_0(1,2...A)$ is given by (\ref{Psi_0}) and
\begin{equation}
\Psi_h(1,2...A-1)= G(1,2...A-1)\Phi_h(1,2...A-1),
\label{Psih}
\end{equation}
where  $\Phi_h(1,2...A-1)$ is a Slater determinant obtained 
by removing from $\Phi_0(1,2...A)$ a nucleon  in the state $h$. 

In order to develop a cluster expansion for $\psi_h(r)$ it is 
convenient to rearrange (\ref{QH}) as:
 \begin{equation}
\psi_h(r) = {\cal X}_h(r) {\cal N}_h^{1/2}
\, ,
\label{QH1}
 \end{equation}
where
 \begin{equation}
 {\cal X}_h(r)= 
{\sqrt A} {
{\langle \Psi_h(A-1)\vert 
{\cal Y}_{lm\sigma \tau}({\hat r}, {\bf \sigma}, {\bf \tau})
\delta ({\bf r}-{\bf r}_A) \vert \Psi_0(A)\rangle }
\over
{\langle \Psi_h(A-1)\vert \Psi_h(A-1)\rangle }}
\, ,
\label{QH2}
 \end{equation}
and
 \begin{equation}
{\cal N}_h=
 {
{\langle \Psi_h(A-1)\vert \Psi_h(A-1)\rangle}
\over
{\langle \Psi_0(A)\vert \Psi_0(A)\rangle} }
 \, .
\label{QH3}
 \end{equation}

Cluster expansions are used to compute ${\cal X}_h$ and ${\cal N}_h$, 
along the lines followed in Ref.\cite{ben89} to evaluate the overlap 
matrix elements in the CBF approach to the nuclear matter 
spectral function.  

The expansion for ${\cal X}_h$ is linked, in the sense 
that disconnected diagrams in the numerator are exactly canceled by those 
coming from the denominator. Its FHNC/0 expression, when only Jastrow 
correlations are considered, is: 
\begin{eqnarray}
 {\cal X}^J_{nl}(r) &=&
\xi^{nl}_\omega(r)
\left \{ 
R_{nl}(r) + \int d^3 r_1 R_{nl} (r_1) 
P_l(\cos \theta) 
\left [
g^{nl}_{\omega d}({\bf r},{\bf r}_1)
C^{nl}_d({\bf r}_1)
\right . 
\right . 
 \\ \nonumber &\times&   
\left . 
\left .
\left (
-\rho_{IPM}^{nl}({\bf r},{\bf r}_1)
+N_{\omega_c c}^{nl}({\bf r},{\bf r}_1)
\right )
+\rho_{IPM}^{nl}({\bf r},{\bf r}_1)
-N_{\omega_c \rho}^{nl}({\bf r},{\bf r}_1)
-N_{\rho \rho}^{nl}({\bf r},{\bf r}_1)
\right ]
\right \}
 \, ,
\label{Xh_J}
\end{eqnarray}
where $\theta$ is the angle betwen ${\bf r}$ and ${\bf r}_1$, and 
$\rho_{IPM}^{h}$ is the A-1 one--body 
density matrix in the independent particle model, 
\begin{equation}
\rho_{IPM}^h({\bf r}_1,{\bf r}_{1'})=
\sum_{\alpha\neq h} \phi_\alpha^\dagger(1) \phi_\alpha(1')
\, .
\label{OBDM_h}
\end{equation}

In ${\cal N}_h$ only those diagrams from the denominator containing 
explicitely the $h$--orbital survive. The Jastrow, FHNC/0 expression is:
\begin{eqnarray}
 [{\cal N}^J_{nl}]^{-1} &=& 
\int d^3 r C^{nl}_d({\bf r})
\left \{ \vert \phi_{nl}({\bf r})\vert ^2
 + \int d^3 r_1 \phi_{nl}^\dagger ({\bf r})\phi_{nl} ({\bf r}_1)
\left [
g^{nl}_{dd}({\bf r},{\bf r}_1)
C^{nl}_d({\bf r}_1)
\right . 
\right . 
 \\ \nonumber &\times&   
\left . 
\left .
\left (
-\rho_{IPM}^{nl}({\bf r},{\bf r}_1)
+N_{cc}^{nl}({\bf r},{\bf r}_1)
\right )
+\rho_{IPM}^{nl}({\bf r},{\bf r}_1)
-N_{x \rho}^{nl}({\bf r},{\bf r}_1)
-N_{\rho \rho}^{nl}({\bf r},{\bf r}_1)
\right ]
\right \}
 \, .
\label{Nh_J}
\end{eqnarray}
The FHNC quantities entering  
${\cal X}^J_{nl}$ and ${\cal N}^J_{nl}$ 
correspond to  those given in Refs.\cite{co94,co92}, but 
 evaluated with the A-1 densities, 
$\rho_{IPM}^h({\bf r}_1,{\bf r}_{1'})$ and 
$\rho_{1,IPM}^h({\bf r}_1)=\sum_{\alpha\neq h} \vert \phi_\alpha(1)\vert ^2$. 
In absence of correlations,  
${\cal X}^J_{nl}(r)\rightarrow R_{nl}(r) $ and 
${\cal N}^J_{nl}\rightarrow 1 $.

The FHNC/SOC expressions of ${\cal X}_{nl}$ and ${\cal N}_{nl}$,  
with operatorial correlations, are given in the Appendix.

The quasihole normalization gives the spectroscopic factor, $S_h$, 
\begin{equation}
S_h=\int r^2 dr \psi_h^2(r)
\, .
\label{factor}
\end{equation}
In a fixed center IPM (as the one we adopt as model function), 
$S_h^{IPM}=1$. Center of mass ($cm$) corrections are  sources 
of deviation. In the harmonic oscillator model they enhance $S_h$ 
for the valence hole states (those with the largest oscillator 
quantum number, $N_v$) by a $[A/(A-1)]^{N_v}$ factor\cite{die74}. 
As a consequence, the $cm$--corrected $1p$--shell 
spectroscopic factor of $^{16}$O is $S_{1p,cm}^{HO}=16/15\sim 1.07$, 
while the average between the $2s$ and $1d$ states 
in $^{40}$Ca is $S_{2s/1d,cm}^{HO}=(40/39)^2\sim 1.05$. More realistic 
Woods--Saxon orbitals do not allow for an analytical treatment of 
$cm$ effects, which have to be computed numerically. It has been found that 
in $^{16}$O the $1p$ WS spectroscopic factor practically coincides with 
the HO one\cite{nec98}.

The correlated spectroscopic factors (without $cm$ corrections) in 
$^{16}$O and $^{40}$Ca are given in Table \ref{tab:tab3} for 
$f_6$, $f_4$ and Jastrow correlation factors. Jastrow correlations 
marginally reduce $S_h$ (at most 3$\%$).
%
%
The Jastrow  $l$--th spectroscopic factors may result slightly larger 
than the total occupation of the corresponding natural orbits, 
given in Table \ref{tab:tab2}. A similar feature was found in 
Ref.\cite{van97}. The small deviations of the Jastrow 
model from the correct behavior are, in our opinion, well within 
the accuracy of the numerical procedures we have adopted and of the 
approximations in the cluster summation.
%
%
 Central spin--isospin 
correlations ($f_4$ model) also provide a few percent depletion in 
the valence states, whereas the tensor ones ($f_6$) give most of the 
reduction bringing $S_{1p}$ in $^{16}$O to 0.90 and 
$S_{2s}$ and $S_{1d}$ in $^{40}$Ca to 0.86 and 0.87, respectively. 
The $1p$ CBF $^{16}$O result is in complete agreement with the VMC 
estimate\cite{nec98}. The influence of the operatorial correlations 
is much larger in the low lying states, whose spectroscopic factors 
are drastically reduced by both central and tensor components:
$S_{1s}$ in $^{16}$O is 0.70, 
$S_{1p}$ and $S_{1s}$ in $^{40}$Ca are 0.58 and 0.55, respectively. 
An analogous behavior was found by Benhar\cite{ben86}, who first used 
low order cluster expansions to estimate $S_h$ in the 
$^{12}$C$(e,e'p)^{11}$B reaction with state dependent correlations 
and found $S_{1p}$=0.55 and $S_{1s}$=0.25.
 
 Results similar to those presented in this paper have been obtained 
in Refs.\cite{gai99,nec97}, where the $^{16}$O $S_{1p}$ has been 
extracted by several models of OBDM\cite{nec93}. In particular, in 
both references it is confirmed that correlation effects on this  
spectroscopic factor are dominated by tensor components. The lowest
order truncation of the OBDM cluster expansion adopted in 
Ref.\cite{nec97} provides $S_{1p,LO}\sim$ 0.90, in agreement with 
the FHNC/SOC results. However, the authors find $S_{1s,LO}\sim$ 0.86, 
in contrast with the 0.70 FHNC/SOC value. The origin of this large 
difference may lay in the  lowest order approximation in the cluster 
expansion. This issue is presently object of investigation.

The latest experimental extraction of $S_p$ from the 
$^{16}$O$(e,e'p)^{15}$N reaction\cite{leu94} reports 
$S_{p_{1/2}}$=0.61 for the 1/2$^{-}$ ground state in $^{15}$N and  
$S_{p_{3/2}}(6.32)$=0.53 for the lowest 3/2$^{-}$  state at 6.32 MeV. This 
state exhibits 87$\%$ of the total $S_{p_{3/2}}$ strength, that is 
fragmented over three states at 6.32, 9.93 and 10.70 MeV. So, the total  
$S_{p_{3/2}}$ may be estimated to be 
$S_{p_{3/2}}$=0.53/0.87=0.61\cite{nec98}. 

A corresponding situation is met in the $^{40}$Ca$(e,e'p)^{39}$K 
reaction\cite{lap93}, where the transition to the 1$d_{3/2}$ ground 
state gives  $S_{d_{3/2}}\sim$0.61$\pm$0.07, while the FHNC/SOC value 
is $S_{d}=$0.87. The $^{40}$Ca spectroscopic factors have been computed 
by the low order cluster expansion of the OBDM in a Jastrow model 
in Ref.\cite{sto96}. 
The results are consistenly lower than the FHNC ones, reported in the 
5th column of Table \ref{tab:tab3}. For instance, 
$S_{2s,LO}$=0.95 and $S_{1d,LO}$=0.91, whereas 
$S_{2s,FHNC}$=0.98 and $S_{1d,FHNC}$=0.97. The discrepancies are 
probably to be ascribed to the approximation used in the reference 
to evaluate the OBDM. 

\begin{figure}
\epsfig{file=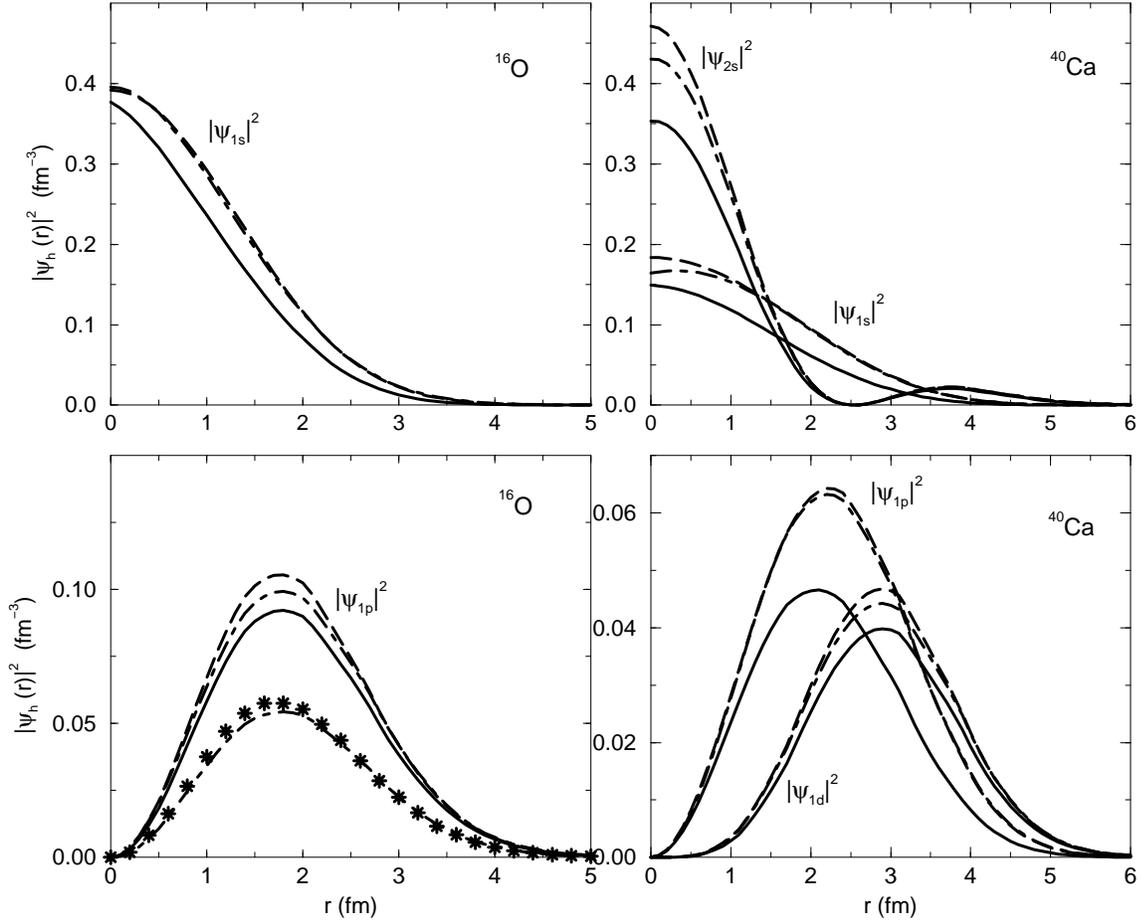,width=15 cm}
\caption{
Squared quasihole wave functions. 
Solid lines: $f_6$ model; dot--dashed: Jastrow; dashed: IPM. 
The $1p$ panel of $^{16}$O shows also the empirical overlap 
(stars) and the $f_6$ one, rescaled as explained 
in the text (lower dot--dashed line)
}
\label{fig:fig7}
\end{figure}

The squared quasihole functions are shown in 
 Fig.\ref{fig:fig7}. The solid and dot--dashed lines give the full $f_6$ and 
Jastrow results, respectively. The IPM estimates are given as dashed lines. 
 The spin--isospin dependent correlations are the main responsible 
for the quenching of the IPM QH functions
%
%
 and, consequently, of the spectroscopic factor. The 
%
%
Jastrow components have little effect on the overlaps, 
and mostly in the valence states. In Ref.\cite{rad94} a Woods--Saxon 
potential was used to generate a single--particle wave function to fit 
the $^{16}$O$(e,e'p)^{15}$N cross section to the 6.32 MeV state with 
$S_{p_{3/2}}(6.32)$=0.53. $\vert\psi_{WS}\vert^2$ is shown in the 
$\vert\psi_{1p}\vert^2$--$^{16}$O panel as stars. In order to give 
a meaningful comparison, we rescale $\vert\psi_{1p,FHNC}\vert^2$ by 
the factor 0.53/0.90. The result is shown as a dot--dashed line and 
it is in nice agreement with the empirical estimate. 

\begin{figure}
\epsfig{file=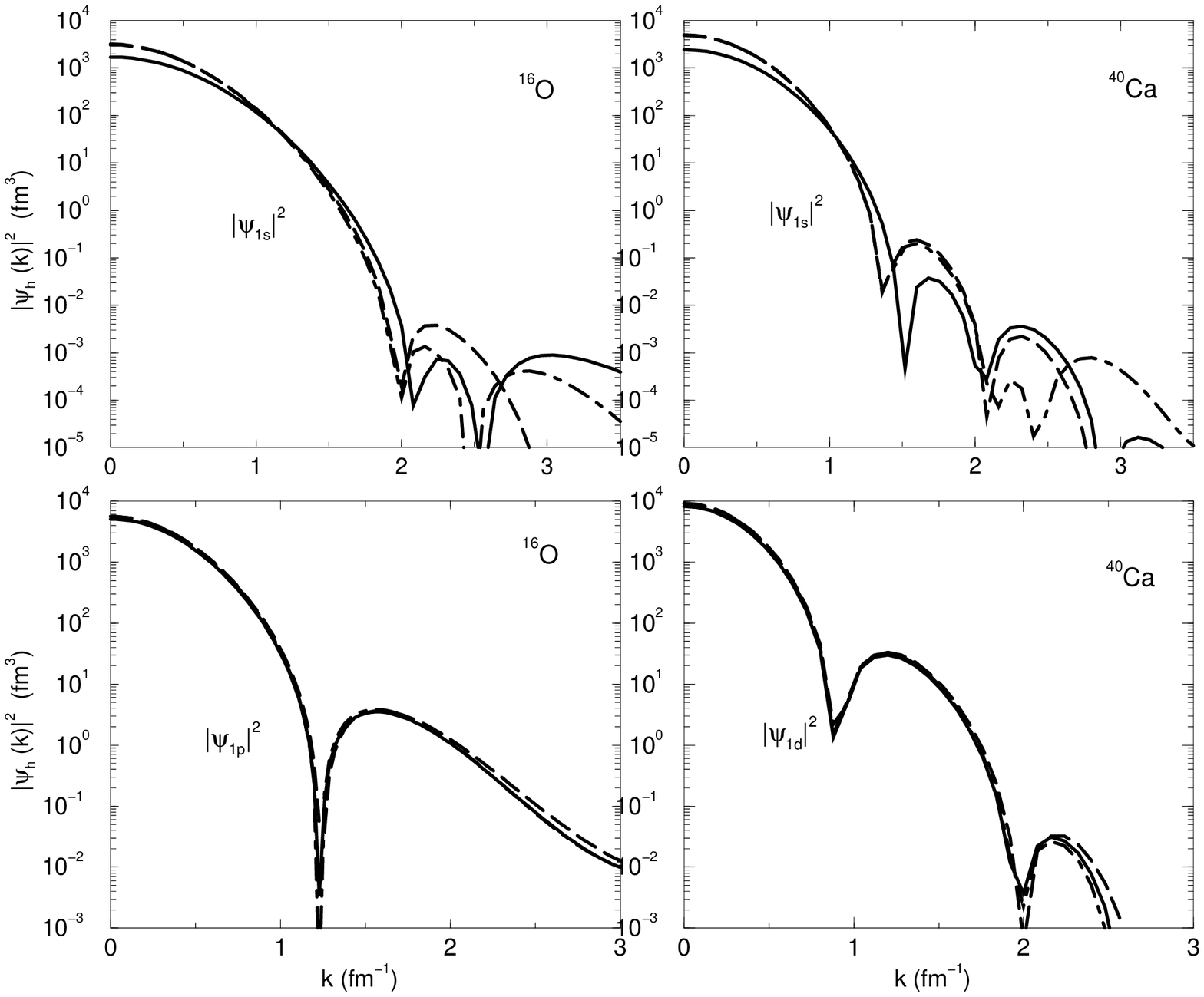,width=15 cm}
\caption{
Squared transform of the quasihole wave function for 
$^{16}$O and $^{40}$Ca. 
Solid lines: $f_6$ model; dot--dashed: Jastrow; dashed: IPM. 
}
\label{fig:fig8}
\end{figure}

The knowledge of $\psi_h(r)$  may give access to the 
cross sections. However, both Coulomb distortion and 
final state interactions should be properly accounted for, by 
evaluating the Fourier transorm of a distorted overlap\cite{ben00}, to 
perform a quantitative comparison with the experiments. 
Work in this direction is in progress. In this paper we limit 
ourselves to give in Fig.\ref{fig:fig8} the squared Fourier transform of 
some quasihole functions,
\begin{equation}
\psi_h(k)=\int d^3r e^{i {\bf k}\cdot {\bf r}} \,\,\psi_h(r)
\, .
\label{transform}
\end{equation}

In the valence states ($1p$ for $^{16}$O and $1d$ for $^{40}$Ca)  
short range correlations slightly deplete $\vert\psi_h(k)\vert^2$ at large 
momenta with respect to the IPM. 
%
%
 This behavior is in contrast with that of the total momentum distributions 
at large--$k$, given in Fig.\ref{fig:fig2} and showing a large enhancement 
due to the correlations. This discrepancy has been already 
observed\cite{van97} and it is confirmed by our approach. 
%
%
 The effect of the correlations is 
more visible in the two low lying states given in the figure 
($1s$ for both nuclei). For instance, in the $1s$ $^{16}$O case 
Jastrow correlations are effective at large momenta only, beyond 
the first IPM zero; instead, tensor correlations modify both the low 
and large momenta behaviors. The same effect is found in $^{40}$Ca. 

\section{Summary and conclusions}

In this work we have calculated one-body density matrices, momentum
distributions, natural orbits and quasi-hole states of $^{16}$O and
$^{40}$Ca using the FHNC/SOC resummation technique, which allows for 
using realistic interactions and state dependent correlations. The
calculations have been done with the Argonne $v_8'$ two--nucleon potential 
plus the Urbana IX three--nucleon interaction, together with a set of 
single particle wave functions fixed to reproduce at best the empirical charge
distributions of the two nuclei. The parameters of the correlation 
have been chosen to minimize the binding energies. Using these wave 
functions, we have investigated the role of
the correlations on the quantitites above mentioned.

Our density and momentum distributions confirm some well known
results. Short-range correlations have small effects on the density
distributions and mainly around the center of the nucleus. On the
contrary, the high momentum tail of the momentum distribution is
dominated by the correlations. We have pointed out that the tensor
correlations enhance these tail by a factor 3-4 with respect to the
results obtained with Jastrow correlations.

The tensor part of the correlation is important in the calculation of
the occupation probabilities of the natural orbits. The effect of
reducing the occupation of the level below the Fermi surface and
enhancing those which lies above is amplified by the tensor terms of
the correlation. We found that the shape of the natural orbits below
the Fermi surface is rather similar to that of the corresponding
single particle wave functions.

The natural orbits have also beeen calculated within a lowest-order
computational scheme. The agreement with the orbit below the Fermi
surface is excellent. The lowest order calculation produces orbits
above the Fermi surface with completely different shape with respect
to those obtained by the full calculation.

Tensor correlations play a relevant role also for the evaluation of
the overlap functions and spectroscopic factors. The correlated overlap 
functions are close to the corresponding single particle wave functions 
if only Jastrow correlations are used. The inclusion of the tensor 
correlations strongly modifies their shapes. 
This behavior is also clear from the analysis of the spectroscopic factors. 
The depletion of a few percent, with respect to one, obtained with 
Jastrow correlations, becomes of about 10-15\%
for the valence levels and 30-45\% for the deeply lying ones. 

In spite of this noticeable reduction, the FHNC/SOC approach in $^{16}$O 
is still unable to reproduce the empirical $S_{p_{3/2}}$ spectroscopic 
factor extracted from $(e,e'p)$ reactions. A similar behavior was found 
in Ref.\cite{ben90} for nuclear matter, where the variational FHNC/SOC 
calculation of the one--hole strength, $Z(e)$, around the Fermi level 
provided $Z_v(e\sim e_F)\sim$0.88,  mostly due to tensor correlations. 
Second order perturbative corrections in a correlated basis, obtained 
by considering the contribution of two--hole one--particle, $(2h-1p)$,  
correlated states, $\Psi_{2h-1p}= G\Phi_{2h-1p}$, were found to 
bring the strength to $Z_{CBF}(e\sim e_F)\sim 0.70$, so 
explaining almost half of the discrepancy with the empirical $^{208}$Pb 
spectroscopic factor, $Z(^{208}Pb)\sim 0.5-0.6$. 
The remaining part of the difference was attributed to the coupling of
the single particle waves to the collective low-lying surface
vibrations, not reproducible in infinite nuclear matter.  
We expect that in finite nuclei, the inclusion of correlated 
$2h-1p$ corrections can take into account also great part of 
the coupling with surface vibrations.

\section*{Aknowledgments}
This work has been partially supported by MURST through the  
{\sl Progetto di Ricerca di Interesse Nazionale: 
Fisica teorica del nucleo atomico e dei sistemi a molticorpi}.

\section*{Appendix}

We present in this Appendix the FHNC/SOC expressions for the ${\cal X}_{nl}$ 
and ${\cal N}_{nl}$ functions:
\begin{eqnarray}
 {\cal X}_{nl}(r) &=&
 {\cal X}^J_{nl}(r) 
 \left [ 1 + \Delta \xi^{op,nl}_\omega(r) \right ]
  +  \xi^{nl}_\omega(r) 
 \int d^3 r_1 R_{nl} (r_1) 
P_l({\hat r}, {\hat r}_1) 
 \\ \nonumber &\times& 
\left \{  
g^{c,nl}_{\omega d}({\bf r},{\bf r}_1)
C^{nl}_d({\bf r}_1) U^{op,nl}_d({\bf r}_1)
\left (
-\rho_{IPM}^{nl}({\bf r},{\bf r}_1)
+N_{\omega_c c}^{c,nl}({\bf r},{\bf r}_1)
\right )  
\right .
 \\ \nonumber &+&   
\sum_{p\geq 2} \left [
h^{p,nl}_{\omega }({\bf r},{\bf r}_1)
h^{c,nl}_{\omega }({\bf r},{\bf r}_1)
C^{nl}_d({\bf r}_1)
\left (
-\rho_{IPM}^{nl}({\bf r},{\bf r}_1)
+N_{\omega_c c}^{c,nl}({\bf r},{\bf r}_1)
\right )  
\right .
 \\ \nonumber &+&   
\left .
\left .
g^{c,nl}_{\omega d}({\bf r},{\bf r}_1)
C^{nl}_d({\bf r}_1)
N_{\omega_c c}^{p,nl}({\bf r},{\bf r}_1)
-N_{\omega_c \rho}^{p,nl}({\bf r},{\bf r}_1)
-N_{\rho \rho}^{p,nl}({\bf r},{\bf r}_1)
\right ] A^p \Delta^p
\right \} 
 \, ,
\label{Xh_op}
\end{eqnarray}
\begin{eqnarray}
 [{\cal N}_{nl}]^{-1} &=&
\int d^3 r C^{nl}_d({\bf r})
\left [
 1 +  U^{op,nl}_d({\bf r}) 
\right ]
\left \{
\vert \phi_{nl}({\bf r})\vert ^2
+ \int d^3 r_1 \phi_{nl}^\dagger ({\bf r})\phi_{nl} ({\bf r}_1)
\right . 
 \\ \nonumber &\times&   
\left [
g^{nl}_{dd}({\bf r},{\bf r}_1)
C^{nl}_d({\bf r}_1)
\left (
-\rho_{IPM}^{nl}({\bf r},{\bf r}_1)
+N_{cc}^{nl}({\bf r},{\bf r}_1)
\right )
\right . 
 \\ \nonumber &+&
\left .
\left .
\rho_{IPM}^{nl}({\bf r},{\bf r}_1)
-N_{x \rho}^{nl}({\bf r},{\bf r}_1)
-N_{\rho \rho}^{nl}({\bf r},{\bf r}_1)
\right ]
\right \}
 \\ \nonumber &+&
\int d^3 r C^{nl}_d({\bf r}) \int d^3 r_1 
\phi_{nl}^\dagger ({\bf r})\phi_{nl} ({\bf r}_1)
 \\ \nonumber &\times&   
\left \{  
g^{c,nl}_{dd}({\bf r},{\bf r}_1)
C^{nl}_d({\bf r}_1) U^{op,nl}_d({\bf r}_1)
\left (
-\rho_{IPM}^{nl}({\bf r},{\bf r}_1)
+N_{cc}^{c,nl}({\bf r},{\bf r}_1)
\right )  
\right .
 \\ \nonumber &+&   
\sum_{p\geq 2} 
\left [
h^{p,nl}({\bf r},{\bf r}_1)
h^{c,nl}({\bf r},{\bf r}_1)
C^{nl}_d({\bf r}_1)
\left (
-\rho_{IPM}^{nl}({\bf r},{\bf r}_1)
+N_{c c}^{c,nl}({\bf r},{\bf r}_1)
\right )  
\right .
 \\ \nonumber &+&   
\left .
\left .
g^{c,nl}_{dd}({\bf r},{\bf r}_1)
C^{nl}_d({\bf r}_1)
N_{cc}^{p,nl}({\bf r},{\bf r}_1)
-N_{x \rho}^{p,nl}({\bf r},{\bf r}_1)
-N_{\rho \rho}^{p,nl}({\bf r},{\bf r}_1)
\right ] 
A^p \Delta^p
\right \} 
 \, .
\label{Nh_op}
\end{eqnarray}
%

%
%
%
%
\newpage
%
%
\begin{table}
\caption{
Occupation numbers of the $nl$--th natural 
orbits for $^{16}$O and $^{40}$Ca in CBF, with the $f_6$ and 
Jastrow correlation models.
} 
\begin{tabular}{ccccc}
    $nl$ 
  &$n_{nl}$($f_6$;$^{16}$O) & $n_{nl}$(J;$^{16}$O)  
  &$n_{nl}$($f_6$;$^{40}$Ca)& $n_{nl}$(J;$^{40}$Ca) 
 \\
\tableline
 $1s$ & 0.858 & 0.960 & 0.864 & 0.952 \\
 $2s$ & 0.019 & 0.005 & 0.780 & 0.962 \\
 $3s$ & 0.010 & 0.002 & 0.052 & 0.002 \\
 $4s$ & 0.005 & 0.001 & 0.013 & 0.001 \\
 $1p$ & 0.919 & 0.980 & 0.841 & 0.949 \\
 $2p$ & 0.021 & 0.004 & 0.024 & 0.009 \\
 $3p$ & 0.011 & 0.003 & 0.016 & 0.006 \\
 $1d$ & 0.025 & 0.006 & 0.956 & 0.983 \\
 $2d$ & 0.011 & 0.003 & 0.030 & 0.007 \\
 $3d$ & 0.006 & 0.001 & 0.019 & 0.006 \\
\end{tabular}
\label{tab:tab1}
\end{table}
\newpage
%
%
\begin{table}
\caption{
The occupation numbers of the $l$--th natural 
orbits for $^{16}$O and $^{40}$Ca in CBF, with three correlation 
models and in the Green function (GF) approach.
} 
\begin{tabular}{cccccccccc}
  & corr. 
  &$n_s      $ &$n_p      $ &$n_d      $ &$n_f      $ 
 \\
\tableline
 $^{16}$O &   J   & 0.971 & 0.991 & 0.011 & 0.002 
 \\
          & $f_4$ & 0.977 & 0.988 & 0.015 & 0.004 
 \\
          & $f_6$ & 0.899 & 0.966 & 0.047 & 0.006 
 \\
          &  GF   & 0.936 & 0.951 & 0.022 & 0.007 
 
 \\
 $^{40}$Ca&   J   & 1.932 & 0.975 & 1.003 & 0.022 
 \\
          & $f_4$ & 1.907 & 0.964 & 1.005 & 0.022 
 \\
          & $f_6$ & 1.727 & 0.920 & 1.026 & 0.071 
\end{tabular}
\label{tab:tab2}
\end{table}
\newpage
%
%
\begin{table}
\caption{
CBF spectroscopic factors for $^{16}$O and $^{40}$Ca, with Jastrow (J) and 
spin--isospin correlations, with ($f_6$) and without ($f_4$) 
tensor components.
} 
\begin{tabular}{cccccccccc}
  & corr. &$1s$ &$1p$ &$1d$ &$2s$ 
 \\
\tableline
 $^{16}$O &   J   & 0.98 & 0.98 &      &      
 \\
          & $f_4$ & 0.79 & 0.96 &      &      
 \\
          & $f_6$ & 0.70 & 0.90 &      &      
 \\
 $^{40}$Ca&   J   & 0.98 & 0.99 & 0.97 & 0.98 
 \\
          & $f_4$ & 0.71 & 0.76 & 0.96 & 0.97 
 \\
          & $f_6$ & 0.55 & 0.58 & 0.87 & 0.86 
\end{tabular}
\label{tab:tab3}
\end{table}

\newpage
%
%


\begin{references}
\bibitem{zab78} 
J. G. Zabolitzky and W, Ey, Phys. Lett. B {\bf 76}, 527 (1978);
O. Bohigas and S. Stringari, Phys. Lett. B {\bf 95}, 9 (1980);
M. Dal R\`{\i}, S. Stringari and O. Bohigas,
Nucl. Phys. A {\bf 376}, 81 (1982);
F. Dellagiacoma, G. Orlandini and M. Traini,  
Nucl. Phys. A {\bf 393}, 95 (1983);
M. Jaminon, C, Mahaux and H. Ng\^o, Nucl. Phys. A {\bf 473}, 509
(1987).
\bibitem{ant88} A. N. Antonov, P. E. Hodgson and I. Zh. Petkov
{\em Nucleon momentum and density distributions} 
(Clarendon Press, Oxford, 1988).
\bibitem{ste91} G. van der Steenhoven and P. K. A. de Witt Huberts,
 in {\em Modern Topics in Electron Scattering}, edited by B. Frois 
 and I. Sick (World Scientific, Singapore, 1991), p.510.
\bibitem{wir00} R. B. Wiringa, S. C. Pieper, J. C. Carlson, 
 and V. R. Pandharipande,  nucl-th/0002022.  
\bibitem{pud97} B. S. Pudliner, V. R. Pandharipande, J. Carlson, S. C. Pieper, 
 and R. B. Wiringa, 
 Phys. Rev. C {\bf 56}, 1720 (1997).
\bibitem{che86} C. R. Chen, G. L. Payne, J. L. Friar, and B. F. Gibson, 
 Phys. Rev. C {\bf 33}, 1740 (1986); 
 A. Stadler, W. Gl\"ockle, and P. U. Sauer, Phys. Rev. C {\bf 44}, 2319 (1991).
\bibitem{kie93} A. Kievsky, M. Viviani, and S. Rosati, 
 Nucl. Phys. A {\bf 551}, 241 (1993).
\bibitem{pie92} S. C. Pieper, R. B. Wiringa, and V. R. Pandharipande, 
 Phys. Rev. C {\bf 46}, 1741 (1992).
\bibitem{hei99} J. H. Heisenberg and B. Mihaila, 
 Phys. Rev. C {\bf 59}, 1440 (1999).
\bibitem{wir88} R. B. Wiringa, V. Ficks, and A. Fabrocini, 
 Phys. Rev. C {\bf 38}, 1010 (1988).
\bibitem{co92} G. Co', A. Fabrocini, S. Fantoni, and I. E. Lagaris, 
 Nucl. Phys. A {\bf 549}, 439 (1992).
\bibitem{ari96} F. Arias de Saavedra, G. Co', A. Fabrocini, and S. Fantoni, 
 Nucl. Phys. A {\bf 605}, 359 (1996).
\bibitem{fab00} A. Fabrocini,  F. Arias de Saavedra, and  G. Co',  
 Phys. Rev. C {\bf 61}, 044302 (2000).
\bibitem{pan79} V. R. Pandharipande and R. B. Wiringa, 
 Rev. Mod. Phys.  {\bf 51}, 821 (1979).
\bibitem{fab98} A. Fabrocini,  F. Arias de Saavedra, G. Co', and 
 P. Folgarait, Phys. Rev. C {\bf 57}, 1668 (1998).
\bibitem{ond97} C. J. G. Onderwater et al. Phys. Rev. Lett. {\bf 78},
  4893 (1997);  Phys. Rev. Lett. {\bf 81},
  2213 (1998);  R. Starink et al. Phys. Lett. B {\bf 474}, 33 (2000). 
\bibitem{dew90} P. K. A. de Witt Huberts, J. Phys. G {\bf 16}, 507 (1990).
\bibitem{hyd84} 
C.E. Hyde-Wright {\it et al.},Phys. Rev. C {\bf 35}, 880 (1987);
J. E. Wise {\it et al.}, Phys. Rev. C {\bf 31}, 1699 (1985);
J.P. Connelly {\it et al.}, Phys. Rev. C {\bf 45}, 2711 (1992). 
\bibitem{cav82} J.M. Cavedon {\it et al.}, Phys. Rev. Lett. {\bf 49}, 978
  (1982). 
\bibitem{pap86} C. Papanicolas, in {\sl Nuclear Structure at high spin,
  excitation and momentum transfer}, H. Nann ed., 
 (American Institute of Physics, New York 1986), p. 110.
\bibitem{mau91} C. Mahaux and R. Sartor, Adv. Nucl. Phys. {\bf 20}, 1 (1991). 
\bibitem{kel96} J. J. Kelly, Adv. Nucl. Phys. {\bf 23}, 75 (1996). 
\bibitem{lap93} L. Lapik\'as, Nucl. Phys. A {\bf 553}, 297c (1993).
\bibitem{fan78} S. Fantoni, 
 Nuovo Cimento A {\bf 44}, 178 (1978).
\bibitem{fan83} S. Fantoni and V. R. Pandharipande, 
 Nucl. Phys. A {\bf 427}, 59 (1983).
\bibitem{ben92} O. Benhar, A. Fabrocini, and S. Fantoni, 
 Nucl. Phys. A {\bf 550}, 201 (1992).
\bibitem{co94} G. Co', A. Fabrocini, and S. Fantoni, 
 Nucl. Phys. A {\bf 568}, 73 (1994).
\bibitem{ben86} O. Benhar, C. Ciofi degli Atti, S. Liuti, and G. Salm\`e, 
 Phys. Lett. B {\bf 177}, 135 (1986).
\bibitem{ari97} F. Arias de Saavedra, G. Co', and M.M. Renis,
 Phys. Rev. C {\bf 55}, 673 (1997).
\bibitem{mou00} Ch. E. Moustakidis, and S. E. Massen,
 Phys. Rev. C {\bf 62}, 034318 (2000).
\bibitem{str90} S. Stringari, M. Traini, and O. Bohigas, 
 Nucl. Phys. A {\bf 516}, 33 (1990).
\bibitem{pol95} A. Polls, H. M\"uther, and W. H. Dickhoff, 
 Nucl. Phys. A {\bf 594}, 117 (1995).
\bibitem{lap99} L. Lapik\'as, J. Wesseling, and R.B. Wiringa, 
 Phys. Rev. Lett. {\bf 82}, 4404 (1999).
\bibitem{nec98} D. Van Neck, M. Waroquier, A. E. L. Dieperink, S. C. Pieper, 
 and V. R. Pandharipande,
 Phys. Rev. C {\bf 57}, 2308 (1998).
\bibitem{nec93} D. Van Neck, M. Waroquier, and K. Heyde, 
 Phys. Lett. B {\bf 314}, 255 (1993).
\bibitem{gai99} M. K. Gaidarov, K. A. Pavlova, A. N. Antonov, M. V. Stoitsov, 
 S. S. Dimitrova, M. V. Ivanov, and C. Giusti,
 Phys. Rev. C {\bf 61}, 014306 (1999).
\bibitem{nec97} D. Van Neck, L. Van Daele, and M. Waroquier,
 Phys. Rev. C {\bf 56}, 1398 (1997).
\bibitem{ami97} K. Amir--Azimi--Nili, J. M. Udias, H. M\"uther, L. D. Skouras,
 and A. Polls, 
 Nucl. Phys. A {\bf 625}, 633 (1997).
\bibitem{arn92} R. A. Arndt, L. D. Roper, R. L. Workman, and 
 M. W. McNaughton, Phys. Rev. D {\bf 45}, 3995 (1992).
\bibitem{sto93} V. G. J. Stoks, R. A. M. Klomp, M. C. M. Rentmeester, and 
 J. J. DeSwart, Phys. Rev. C {\bf 48}, 792 (1993).
\bibitem{wir95} R. B. Wiringa, V. G. J. Stoks, and R. Schiavilla, 
 Phys. Rev. C {\bf 51}, 38 (1995).
\bibitem{fab92} A. Fabrocini, V. R. Pandharipande, Q. N. Usmani,  
 Nuovo Cimento D {\bf 5}, 469 (1992).
\bibitem{viv86} M. Viviani, E. Buend\`{\i}a, A. Fabrocini, and S. Rosati, 
 Nuovo Cimento D {\bf 8}, 561 (1986).
\bibitem{wir84} R. B. Wiringa, R. A. Smith, and T. L. Ainsworth, 
 Phys. Rev. C {\bf 29}, 1207 (1984).
\bibitem{lew88} D. S. Lewart, V. R. Pandharipande, and S. C. Pieper, 
 Phys. Rev. B {\bf 37}, 4950 (1988).
\bibitem{mac89} R. Machleidt, 
 Adv. Nucl. Phys. {\bf 19}, 1 (1989).
\bibitem{stoi93} M. V. Stoitsov, A. N. Antonov and S. S. Dimitrova,
  Z. Phys. A {\bf 345} (1993) 345; 
  Phys. Rev. C {\bf  47} R455 (1993); 
  Phys. Rev. C {\bf 48} 74 (1993).  
\bibitem{van97}D. Van Neck, L. Van Daele, Y. Dewulf and M. Waroquier,
  Phys. Rev. C {\bf 56}, 1398 (1997).
\bibitem{die74} A. E. L. Dieperink, and T. de Forest, 
 Phys. Rev. C {\bf 10}, 543 (1974).
\bibitem{ben89} O. Benhar, A. Fabrocini, and S. Fantoni, 
 Nucl. Phys. A {\bf 505}, 267 (1989).
\bibitem{ben86} O. Benhar, in {\sl Research Program at CEBAF (II). 
 Report of the 1986 Summer Study Group}, CEBAF, 501 (1986).
\bibitem{leu94} M. Leuscher {\it et al.}, 
 Phys. Rev. C {\bf 49}, 955 (1994).
\bibitem{sto96} M. V. Stoitsov, S. S. Dimitrova, and A. N. Antonov, 
 Phys. Rev. C {\bf 53}, 1254 (1996).
\bibitem{rad94} M. Radici, S. Boffi, S. C. Pieper, and V. R. Pandharipande, 
 Phys. Rev. C {\bf 50}, 3010 (1994).
\bibitem{ben00} O. Benhar, N. N. Nikolaev, J. Speth, A. A. Usmani, 
 and B. G. Zakharov, 
 Nucl. Phys. A {\bf 673}, 241 (2000).
\bibitem{ben90} O. Benhar, A. Fabrocini and S. Fantoni, 
 Phys. Rev. C {\bf 41}, R24 (1990).
 
%
\end{references}
\end{document}